\documentclass[]{aa}
\usepackage{natbib}
\usepackage{psfig}
\usepackage{txfonts}
\def\HH{\mbox{H$_2$}}

\def\nH2{\mbox{${\rm n}_\HH}$}

\def\qX{q$_{\rm X} \int \tau_{2-1} {\rm dv}$}
\def\NH2{{\rm N}({\rm H}_2)}
\def\pccc{~{\rm cm}^{-3}} 
 
\def\pcc {~{\rm cm}^{-2}}

\def\Tstar#1 {\mbox{${\rm T}_{\rm #1}^*$}}
\def\Tsub#1 {\mbox{${\rm T}_{\rm #1}$}}
\def\TK  {\Tsub K }

\def\degr{$^{\rm o}$}
\def\p{\mbox{$^+$}}
\def\cotw {\mbox{$^{12}$CO}}
\def\coth {\mbox{$^{13}$CO}}

\def\hcop{\mbox{{HCO\p}}}

\def\cch{\mbox{C$_2$H}}
\def\c3h2{\mbox{C$_3$H$_2$}}
  
\def\hhco{\mbox{H$_2$CO}}

\def\ccchh{\mbox{C$_3$H$_2$}}
 
 \def\R0{R$_0$}

\def\ddeg{{}^\circ\kern-.1em}

\def\kms{\mbox{km\,s$^{-1}$}}
\def\ps{\mbox{s$^{-1}$}}
\def\bll{BL Lac}

\def\E#1 {$10^{#1}$}
\def\E#1 {E{#1}}
\def\P#1,{$\nH2\TK~=~#1\times~10^4\pccc$~K}
\def\ec#1,#2,#3,{#1\,(#2)\E{#3}}

\def\zoph{$\zeta$ Oph}
\def\meth{\mbox{CH$_3$OH}}
\def\H3{\mbox{H$_3$}}
\def\C3N{\mbox{C$_3$N}}
\def\C5N{\mbox{C$_5$N}}

\def\ammon{\mbox{N\H3} }

\def\RH2{\mbox {R$_{\rm G}$}}
\def\fH2{\mbox {f$_{\HH}$}}
\def\FH2{\mbox {F$_{\HH}$}}

\begin{document}

\title{Limits on chemical complexity in diffuse clouds: \\
  search for \meth\ and HC$_5$N absorption \thanks{Based on observations obtained
    with the IRAM Plateau de Bure Interferometer and the NRAO VLA 
     telescope.}}

\titlerunning{Search for \meth\ and HC$_5$N absorption along diffuse/translucent
  sightlines}
\author{H. S. Liszt\inst{1}, J. Pety\inst{2,3}, R. Lucas\inst{2}}
\institute{National Radio Astronomy Observatory,
           520 Edgemont Road,
           Charlottesville, VA,
           USA 22903-2475
\and       Institut de Radioastronomie Millim\'etrique,
           300 Rue de la Piscine,
           F-38406 Saint Martin d'H\`eres,
           France
\and       Obs. de Paris, 
           61 av. de l'Observatoire, 75014, Paris, 
           France}


\date{received \today}
\offprints{H. S. Liszt}
\mail{hliszt@nrao.edu}

\abstract
{An unexpectedly complex polyatomic chemistry exists in diffuse clouds,
  allowing detection of species such as \cch, \ccchh, \hhco, and \ammon, 
  which have relative abundances that are strikingly similar to those inferred
  toward the dark cloud TMC-1.}
{We probe the limits of complexity of diffuse cloud polyatomic chemistry.}
{We used the IRAM Plateau de Bure Interferometer to search for galactic
  absorption from low-lying J=2-1 rotational transitions of A- and
  E-\meth\ near 96.740 GHz and used the VLA to search for the J=8-7 transition 
 of HC$_5$N at 21.3 GHz.}
{Neither \meth\ nor HC$_5$N were detected at column densities well below those of all 
  polyatomics known in diffuse clouds and somewhat below the levels 
 expected from comparison with TMC-1.  The HCN/HC$_5$N ratio is at least 
  3-10 times higher in diffuse gas than toward TMC-1.}
{}


\maketitle{}

\section{Introduction}

 As we have shown in a recent series of papers, local diffuse clouds 
seen in cm-wave and mm-wave absorption against extragalactic background 
sources have an unexpectedly rich and robust polyatomic chemistry
(see \cite{LisLuc+06} and references given there).  At lower 
column densities CO, OH, \hcop, \cch\ and \c3h2\ are 
detected but when N(\hcop) $\ga 10^{12}\pcc$ 
or N(\HH) $\ga 5\times 10^{20}\pcc$,  CS, HCN, \ammon\ and \hhco\  
appear with relative abundances like those inferred toward the 
canonical dark cloud TMC-1 \citep{OhiIrv+92}.

 Some fairly complex species are seen in these absorption studies, 
but the real limits of complexity within this chemistry are not 
known.  Most of our work has been at mm-wavelengths while larger
astrophysically-important species are generally heavier so that
the bulk of their rotational population resides in energy levels 
which are best observed at lower frequencies.

An exception to this general scenario is methanol (\meth), 
many of whose lowest rotational transitions (including the 
ground-state E-type transition) occur near 96740 MHz. These lines were
detected in TMC-1 by \cite{FriHja+88} and the relative abundance
of \meth\ with respect to \hcop\ in TMC-1 is 
N(\meth)/N(\hcop) $\approx 0.25$ \citep{OhiIrv+92}.  
Although the generally-accepted chemical scheme for producing methanol 
in dark gas invokes progressive hydrogenation of \hhco\ on grains 
and might not be expected to be a fertile source of molecules 
in lightly-shielded regions, \hhco\ is widely seen in diffuse clouds 
\citep{Nas90,LisLuc95a,LisLuc+06}.  Furthermore, the environment is
rich in atomic hydrogen in diffuse gas and models have been proposed
in which molecules are hydrogenated on grains and released into
the ambient diffuse gas where high abundances persist for some time
\citep{VitWil+00,PriVit+03}.  Alternatively, material may be cycled
through a dense phase, with persistently high molecular abundances
for quite some time thereafter in a more diffuse state \citep{FalPin+06}.
This being the case, at the suggestion of our colleagues, we 
undertook to search for \meth\ absorption using the IRAM Plateau de 
Bure Interferometer.

 An alternative approach to searching for heavier molecules is simply
to follow them to lower frequencies and, subsequent to the \meth\ 
observations described here,  we realized that the cyanopolyynes 
HC$_3$N  and HC$_5$N should be observable with high sensitivity during the 
VLA-eVLA conversion.  Given the similarity in abundance between so 
many species in TMC-1 and diffuse gas, and the high relative abundances of 
the cyanopolyynes in TMC-1 (where N(HCN):N(HC$_3$N):N(HC$_5$N) = 20:6:3)
it seemed appropriate to search for just those species 
which are the particular hallmark of the chemistry in TMC-1.

Section 2 of this work describes the observations and some aspects of 
the spectroscopy of \meth.  Section 3 describes the cyanopolyyne work
and Sect. 4 presents our upper limits on the \meth\ and HC$_5$N 
abundances  and briefly summarizes our absorption line work to date
as well as the physical conditions under which the diffuse cloud
chemistry operates.

\section{\meth\ observations and data}

\begin{table}
\caption{Background sources observed in \meth\ }
\begin{tabular}{lcccccc}
\hline
Source  & l     &  b    & Date & Flux & Time & $\sigma_{l/c} ^1 $  \\
        & \degr & \degr & 2006          &   Jy &   hours$^2$ & $\times10^{-3}$ \\
\hline
B0355+508 & 150.4 & $-$1.6  & July & 2.8 & 3.5  &  8.5 \\
B0415+379 & 161.7 & $-$8.8  & July & 2.6 & 4.2  & 14.5 \\
B2200+420 & 92.6  & $-$10.4 & May & 2.5 & 8.2 &  6.5 \\
\hline
\end{tabular}
\\
$^1~\sigma_{l/c} =$ rms error in the line/continuum ratio \\
$^2$ Integration time on-source equivalent to using 6 antennas \\
\label{tab:obs}
\end{table}

\subsection{Observed sources and technical details}

The data were acquired at the Plateau de Bure Interferometer in May and
July 2006 with 5 or 6 antennas. Table~\ref{tab:obs} summarizes the observed
sightlines, observing dates, approximate quasar fluxes, integration times 
(the on-source time equivalent to having 6 antennas simultaneously observing), 
and the empirically-determined rms error in 
line/continuum ratio in the final, reduced spectra.

Six correlator bands of 20 MHz were concatenated to cover frequencies from
97600 to 97800 MHz (or a $\sim 150\,\kms$ bandwidth) with a channel spacing
of 39.06 kHz or 0.121 \kms\ and a channel width of 70 kHz.  Two additional
correlator bands of 320 MHz were used to measure the 3~mm continuum over
the 580~MHz instantaneous IF-bandwidth available with this generation of
receivers. The fluxes of the quasar continuum were determined relative to
 the primary flux calibrator used at Plateau de Bure, {\it i.e.} MWC349. The
resulting flux accuracy is $\sim 15\%$.

The data were processed inside the \texttt{GILDAS/CLIC}
software\footnote{See \texttt{http://www.iram.fr/IRAMFR/GILDAS} for more 
information about the \texttt{GILDAS} software.}~\citep{pety05}.  After
a standard RF bandpass calibration, the time-dependent amplitude and 
phase gains were computed per baseline on the continuum data, 
assuming a point source.  Those gains were then
applied to the line data taken simultaneously and spectra were 
computed as a weighted temporal average of the visibility amplitudes.

\subsection{Spectroscopy and observed transitions}

 Rest frequencies for the \meth\ transitions (Table~\ref{tab:line}) were 
taken from the NIST list of recommended rest frequencies, found online at
http://physics.nist.gov/cgi-bin/micro/table5/start.pl. Although the
spectroscopic constants have changed slightly, helpful energy level 
diagrams and related information for \meth\ are given
by \cite{Lees73}, \cite{NagKai+79} and \cite{FriHja+88}; \cite{Lees73}
tabulates line strengths and spontaneous emission coefficients.  As noted
in Table~\ref{tab:line}, we observed several J=2$_{\rm K} - 1_{\rm K}$
transitions of A- and E-type \meth\ around 96740 MHz.  For E-\meth\ the J=0
level of the K=-1 ladder is absent owing to symmetry concerns and the
J=$2_{-1} - 1_{-1}$ transition is actually the ground-state E-\meth\ line.
The fourth column of Table~\ref{tab:line} gives the fraction of all A- or
E-\meth\ which resides in the 1$_{\rm K}$ level of the various transitions
when the rotational populations are in equilibrium with the 2.73 K cosmic
microwave background.  The total column density of \meth\ is the sum of all
A- and E-\meth. 

According to \cite{Lees73}, the transitions observed are all of a-type,
with dipole moment of 0.885 D, leading to the spontaneous emission rates
A$_{21}$ shown in Table~\ref{tab:line}.  From standard formulae, given the
assumed excitation and level populations, we may write for either the X=A
or X=E configurations N(X-\meth) = \qX\ , where
the observed optical depth integral over any of the J=2$_{\rm K} - 1_{\rm
  K}$ lines is expressed in \kms\ and values of q$_{\rm X}$ are given in
the last column of Table~\ref{tab:line}.

\begin{table}
\caption[]{\meth\ spectroscopy and column density}
\begin{tabular}{lcccc}
\hline
Line&$\nu$& A$_{21}$ & f$_{\rm low}^1$ & q$^2$ \\
     &MHz & $10^{-6}$ \ps\     &   & cm$^{-2}/\kms\ $ \\
\hline
A $2_0 - 1_0$ & 96741.38 & 5.5 & 0.472 & $2.39\times10^{13}$ \\
E $2_{-1} - 1_{-1}$ & 96739.39 & 3.3 & 0.678 & $2.77\times10^{13}$ \\
E $2_0 - 1_0$ & 96744.55 & 5.5 & 0.042 & $2.68\times10^{14}$ \\
\hline
\end{tabular}
\\
$^1$  f$_{\rm low}$ is the fraction of A- or E-type \meth\ in the lower, J=1, level \\
of the transition in equilibrium with the 2.73 K background \\
$^2$ N(X-\meth) = \qX\ for either the A- or E-state 
\label{tab:line}
\end{table}

\section{Cyanopolyyne observations and data}

\begin{table}
\caption{Background sources observed in HC$_5$N J=8-7}
\begin{tabular}{lccccc}
\hline
Source  & l     &  b    & Time & $\sigma_{l/c} ^1 $ &  $ \sigma \int \tau dv ^2 $ \\
        & \degr & \degr &  hours & $\times10^{-3}$ & $10^{-3}$ \kms\\
\hline
B0212+735 & 128.9 & 12.0  & 1.1 & 1.5 & 2.2 \\
B0355+508 & 150.4 & $-$1.6 & 1.2  &  1.4 & 2.0 \\
B0415+379 & 161.7 & $-$8.8  & 1.2  & 1.3  & 2.5 \\
B2200+420 & 92.6  & $-$10.4 & 1.1  &  1.6 & 3.2 \\
\hline
\end{tabular}
\\
$^1~\sigma_{l/c} =$ rms error in the line/continuum ratio \\
 $^2 \sigma \int \tau dv  =$ rms error in integrated optical depth \\
\end{table}

\begin{figure*}
\psfig{figure=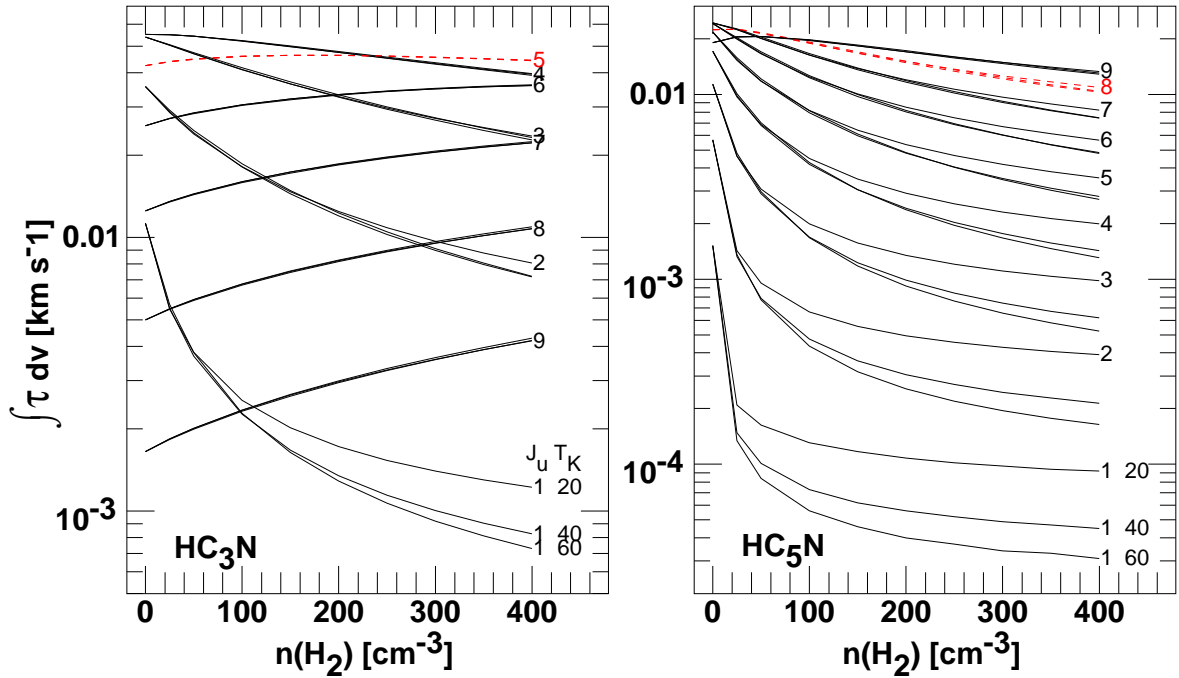,height=9cm}
\caption[]{Excitation of cyanopolyynes by electrons 
in diffuse gas.  Shown are the integrated optical depths
for J$_u \rightarrow {\rm J}_u -1$ transitions as a function of 
\HH-number density when X(e) = n(e)/n(\HH) $= 4\times 10^{-4}$. 
At left are shown results for N(HC$_3$N) = $10^{12} \pcc$
and at right for  N(HC$_5$N) = $10^{12} \pcc$.  For each transition,
calculations at \TK = 20, 40, and 60 K are shown but only for 
lower-J transitions are they clearly separated. In each case the
curve for the highest \TK-value lies lowest.  The transitions
observed in this work are shown as dashed (red) lines.}
\end{figure*}

 We observed the J=5-4 HC$_3$N  and J=8-7 HC$_5$N transitions 
at 45.4 and 21.3 GHz at the VLA on 2007 December 16-17
using a correlator setup with 128 channels of width 24.4 khz 
and 12.2 kHz, respectively (0.161 \kms and 0.172 \kms).
We bandpass calibrated and then observed the background
sources fixedly without the need for other phase calibrators, 
given the strong emission and point-like nature of the sources
(all of which are calibrators for other experiments).
We used reference pointing on all sources.  After applying the
bandpass calibration, we used the AIPS task UVLSD which forms
and averages line/continuum spectra during individual 
correlator integration intervals.   The final spectra were then formed
with vector averaging in the POSSM task and exported for reduction
and analysis.  The fluxes of the background sources were not
needed to form the absorption spectra and were not separately
determined.

 Although unforseen, it has not been possible to correlate
baselines with both VLA and eVLA antennas at the narrow IF 
bandwidths used in this work.  Given the makeup of the VLA during 
our observations, it was necessary to discard nearly half of the
the baselines.  Additionally, the Q-band HC$_3$N observations were 
corrupted by an 
unexplained IF instability or other problem which made passband
calibration problematic and rendered the noise levels several times
higher than expected.  Although some portions of some passbands 
appeared to be usable for some sources, we do not trust these results 
and we will not discuss them further.  Results for the HC$_5$N transition 
toward four sources (those observed in \meth\ as well as 
B0212+735) are summarized in Table 3.

Given the relatively large mass and high dipole moments of the 
cyanopolyynes, 3.6 and 4.33 Debye for HC$_3$N and HC$_5$N, 
respectively, maximizing the sensitivity of the detection 
experiment required consideration of the populations of 
the rotational ladder.  Although the density  of neutral 
particles is too low to produce significant departures from 
rotational equilibrium with the cosmic background  (see Sect. 4.3), 
excitation by electrons is non-negligible .  Fig. 1 shows the 
integrated 
optical depths expected for various rotational transitions 
as a function of the assumed temperature and density of 
molecular hydrogen, under the assumption that
n(e)/n(\HH) $ = 4\times 10^{-4}$  representing a fully ionized
component of moderately-depleted carbon,
n(C) $= 3 \times 10^{-4} n(\HH)$, along with a smaller
contribution by H\p.   We  solved the rate equations determining
the level populations for a total column density 
$10^{12} \pcc$ of absorbers, including collisional excitation 
by electrons using the rate constants of \cite{DicFlo81}.

Technical details aside, 
the transitions having the greatest integrated optical depth
are those which most sensitively probe the actual molecular abundance.
The J=5-4 HC$_3$N and J=8-7 HC$_5$N transitions are the first or
second-most sensitive transitions over the range of density indicated.
The calculated optical depth of the J=8-7 HC$_5$N transition is
insensitive to temperature but declines slightly with increasing
density.  Given the behaviour shown in Fig. 1, it is conservative
to assert that $\int \tau_{8-7} dv = 10^{-14}$ \kms\ N(HC$_5$N) and  
the upper limits on the J=8-7 line profile optical depth integrals 
given in Table 3 have been converted to HC$_5$N column 
density in Table 4 using this value.

\section{Results and comparison with dark cloud abundances}

\subsection{\meth}

The bottom row of Table~\ref{tab:coldens} gives limits on the total \meth\ 
column density toward B0415+479 (3C111) and B2200+420 (\bll) and for the
-10.5 \kms\ component toward B0355+508 (NRAO150), which has the highest
column densities and is chemically the most complex feature along that
sightline \citep{LisLuc+06}.  These are $2\sigma$ statistical upper limits 
at the
empirically-determined channel-to-channel rms levels tabulated in
Table~\ref{tab:obs}, over the expected velocity span determined by our deep
\hcop\ profiles for each line.  The features toward 3C111 and \bll\ have
blended velocity substructure, but this distinction is ignored here.
Spectra of the various species in the directions discussed are given in the
references cited in Table~\ref{tab:coldens}.

Table~\ref{tab:coldens} also compares these limits on the \meth\ column
density with values for the column densities of a variety of molecules
previously observed toward the various features in our earlier work.  To
compare with dark cloud values, the right-most column of
Table~\ref{tab:coldens} gives the abundances of the various species
seen in TMC-1. 

Our upper limits on the \meth\ column density are in all cases quite low
compared to those of the other species shown in Table~\ref{tab:coldens},
and are generally at or modestly below the abundance ratios seen in TMC-1,
especially toward 3C111.  For instance N(\meth)/CS $<$ 0.2, 0.1, and 0.13 
for \bll, NRAO150, and 3C111, respectively, compared with a value 0.2 
toward TMC-1. 

\subsection{HC$_5$N}

As noted in Section 3, the upper limits on the line profile integral
of HC$_5$N absorption in Table 3 were converted to column density 
for inclusion in Table 4 using 
N(HC$_5$N) $= 10^{14} \pcc \int \tau dv$, following the excitation 
calculations shown in Fig. 1.  The HCN/HC$_5$N ratio, approximately
7 in TMC-1,  is at least 3-10 times higher  than this toward 
B2200 and B0415+379.  

\subsection{ Chemical abundances and physical conditions in 
diffuse clouds}


Table~\ref{tab:coldens} serves as a summary of our 
absorption line chemistry work to date, for sightlines and clouds 
with somewhat higher column density N(\hcop) $> 10^{12}\pcc$ 
which have the richest chemistry.  These patterns
are not universal: the abundances of CO and all other detected species 
listed beneath \c3h2\ in the table increase dramatically with respect 
to \hcop\ for N(\hcop) $ \ga 10^{12}$, as shown for instance in 
Fig. 3 of \cite{LisLuc01}.  CO, which is found 
in nearly all features identified in \hcop, even at 
N(\hcop) $ < 10^{12}$, is a special case, varying widely due to 
the influence of photodissocation and self-shielding
\citep{Lis07CO} .  It can however 
be understood as the electron recombination product of \hcop\ when 
N(\hcop)/N(\HH) $= 2\times 10^{-9}$, as observed ({\it ibid}).

Despite the overall similarity in relative abundances of many 
species with the TMC-1 patterns, some differences with TMC-1 are 
also apparent, even beyond the absence of \meth\ and HC$_5$N.  
In particular, the low HNC/HCN ratio in diffuse clouds 
is characteristic of warmer gas, consistent with the observed 
HOC\p/\hcop\ ratio \citep{LisLuc01,LisBla+04}. The HCN/HNC and 
HOC\p/\hcop\ ratios are important clues to the diffuse nature 
of the host gas.
Previous indications that diffuse gas was being observed were
the low reddening (0.32 mag) known to exist toward B2200+420 
(BL Lac),  the weakness of mm-wave emission from species other 
than CO -- only \hcop\ is detected \citep{LisLuc94,LucLis96} --
and finding that N(OH) and N(CO) were comparable to the
column densities observed in $uv$ absorption toward \zoph\ and 
some other bright stars.

The general properties of diffuse gas are  summarized by
\cite{SnoMcC06}.
In the context of our work, the kinetic temperature and the 
density and thermal partial pressure of 
\HH\ are indicated in various ways by the chemistry, 
fractionation and rotational excitation of CO 
\citep{LisLuc98,Lis07CO}, and are typical of the diffuse ISM.  
The partial thermal pressures
n(\HH)\TK $\approx 1-5\times10^3 \pccc$ K are comparable to 
those derived for the bulk of the gas from C I fine-structure 
excitation seen in $uv$ absorption \citep{JenTri01}. 
N(\cotw)/N(\coth) ratios may be as low as 15-20 in clouds
with N(CO) $\la 10^{16}\pcc$, from which it may be inferred 
that the kinetic temperature of lines of sight like those 
summarized in Table 4 is 25 - 50 K, somewhat below the mean
kinetic temperature inferred from obsevation of \HH\ itself
(70-80 K, see \cite{RacSno+02}) but consistent with formation
and rotational excitation of CO at n(\HH) $\approx 100 \pccc$.  
The very weak mm-wave emission of optically-thick \hcop\ 
is consistent with such n(\HH) if
n(e)/n(\HH) $\approx 4\times10^{-4}$ as expectedfor diffuse 
gas in which only a small fraction ($\la 1-5$\%) of the free 
gas-phase carbon resides in CO and the rest is in the form
of C\p.

 Despite the consistency of these arguments, it is the case
that no quiescent ion-molecule chemistry will reproduce the 
observed abundances at such low n(\HH).  Some recent models of the 
diffuse cloud chemistry regard these conditions 
as a general background against which transient processes may 
operate \citep{FalPin+06,SmiPav+04}, affecting the observed chemical abundance 
patterns without necessarily imprinting themselves observably 
on the internal degrees of freedom in the molecules themselves.

\begin{table}
\caption[]{Column densities and relative abundances}
{
\begin{tabular}{lcccc}
\hline
Species & B2200 & B0355 & B0415 & TMC-1$^6$ \\
  &$10^{12}\pcc$ & $10^{12}\pcc$ & $10^{12}\pcc$ & $10^{13}\pcc$ \\
\hline
 OH$^7$  &  68   & 34  & 360  & 300 \\
 CO$^8$  &  1.9$\times10^4$ & 0.5$\times10^4$  & 
 5-8$\times10^4$  & 8$\times10^4$ \\
\hcop$^{1}$   & 2.0     & 1.2  & 11.6  & 8 \\
\cch$^{2}$    & 31     & 23   & 75   & 50-100 \\
\c3h2$^{2}$   & 5.0    & 3.7  & 13.3  & 10 \\
\hhco$^{3}$   & 6.2    & 6.9  & 20.5  & 20 \\
CS$^{4}$      & 2.7    & 4.3  & 10.1  & 10 \\
HCS\p  $^{4}$   &   &  & 0.76  & 0.6 \\

SO$^{4}$ &3.43 & 1.66 & 13.0 & 5 \\
\HH S$^{4}$ &0.52 & 0.76 & 1.66 & $<$ 0.5 \\
NH$_3 ^{3}$  &        & 2.5  & 12.2  & 20 \\
CN$^{5}$     &32.9 & 41 & 158  & 30 \\
HCN$^{5}$     & 4.5    & 3.6  & 24.8  & 20 \\
HNC$^{5}$     &0.74 & 1.1 & 5.54 &  20 \\
HC$_5$N$^{5}$ & $<0.20$ & $<0.25$ & $<0.32$  & 3 \\
\meth   & $<$0.52 &$<$0.43 &$<$1.3  & 2 \\
\hline
\end{tabular}}
\\
$^1$\cite{LucLis96}; $^2$  \cite{LucLis00C2H} \\
$^3$\cite{LisLuc+06}; $^4$ \cite{LucLis02} \\
$^5$\cite{LisLuc01}; $^6$  \cite{OhiIrv+92} \\
$^7$\cite{LisLuc00}; 
$^8$\cite{LisLuc98}; 
\label{tab:coldens}
\end{table}

\begin{acknowledgements}
    IRAM is supported by INSU/CNRS (France), MPG (Germany), 
   and IGN (Spain).  The National Radio Astronomy Observatory is operated 
   by AUI, Inc. under a cooperative agreement with the US National Science 
    Foundation.
  We owe the staff at IRAM (Grenoble) and the Plateau de Bure our thanks for 
  their assistance in taking the data.   We thank the scientific staff at the
  VLA, especially Mark Claussen, for assistance in dealing with data-handling 
  issues during the  VLA/eVLA transition. The referee provided a gentle but
  perceptive report which resulted in great improvement of the text.  
  We thank Maryvonne Gerin for encouraging us to search for \meth.
\end{acknowledgements}

\bibliographystyle{apj}

\begin{thebibliography}{25}
\expandafter\ifx\csname natexlab\endcsname\relax\def\natexlab#1{#1}\fi

\bibitem[{{Dickinson} \& {Flower}(1981)}]{DicFlo81}
{Dickinson}, A.~S. \& {Flower}, D.~R. 1981, Mon. Not. R. Astron. Soc., 196, 297

\bibitem[{{Falgarone} {et~al.}(2006){Falgarone}, {Pineau Des For{\^e}ts},
  {Hily-Blant}, \& {Schilke}}]{FalPin+06}
{Falgarone}, E., {Pineau Des For{\^e}ts}, G., {Hily-Blant}, P., \& {Schilke},
  P. 2006, A\&A, 452, 511

\bibitem[{{Friberg} {et~al.}(1988){Friberg}, {Hjalmarson}, {Madden}, \&
  {Irvine}}]{FriHja+88}
{Friberg}, P., {Hjalmarson}, A., {Madden}, S.~C., \& {Irvine}, W.~M. 1988,
  A\&A, 195, 281

\bibitem[{{Jenkins} \& {Tripp}(2001)}]{JenTri01}
{Jenkins}, E.~B. \& {Tripp}, T.~M. 2001, Astrophys. J., Suppl. Ser., 137, 297

\bibitem[{{Lees}(1973)}]{Lees73}
{Lees}, R.~M. 1973, ApJ, 184, 763

\bibitem[{{Liszt} \& {Lucas}(2001)}]{LisLuc01}
{Liszt}, H. \& {Lucas}, R. 2001, A\&A, 370, 576

\bibitem[{{Liszt} {et~al.}(2004){Liszt}, {Lucas}, \& {Black}}]{LisBla+04}
{Liszt}, H., {Lucas}, R., \& {Black}, J.~H. 2004, A\&A, 428, 117

\bibitem[{{Liszt} {et~al.}(2006){Liszt}, {Lucas}, \& {Pety}}]{LisLuc+06}
{Liszt}, H., {Lucas}, R., \& {Pety}, J. 2006, A\&A, 448, 253

\bibitem[{{Liszt}(2007)}]{Lis07CO}
{Liszt}, H.~S. 2007, A\&A, 476, 291

\bibitem[{{Liszt} \& {Lucas}(1994)}]{LisLuc94}
{Liszt}, H.~S. \& {Lucas}, R. 1994, ApJ, 431, L131

\bibitem[{{Liszt} \& {Lucas}(1995)}]{LisLuc95a}
---. 1995, A\&A, 299, 847

\bibitem[{{Liszt} \& {Lucas}(1998)}]{LisLuc98}
---. 1998, A\&A, 339, 561

\bibitem[{{Liszt} \& {Lucas}(2000)}]{LisLuc00}
---. 2000, A\&A, 355, 333

\bibitem[{{Lucas} \& {Liszt}(1996)}]{LucLis96}
{Lucas}, R. \& {Liszt}, H.~S. 1996, A\&A, 307, 237

\bibitem[{{Lucas} \& {Liszt}(2000)}]{LucLis00C2H}
---. 2000, A\&A, 358, 1069

\bibitem[{{Lucas} \& {Liszt}(2002)}]{LucLis02}
---. 2002, A\&A, 384, 1054

\bibitem[{{Nagai} {et~al.}(1979){Nagai}, {Kaifu}, {Nagane}, \&
  {Akaba}}]{NagKai+79}
{Nagai}, T., {Kaifu}, N., {Nagane}, K., \& {Akaba}, K. 1979, Publ. Astron. Soc.
  Jpn., 31, 317

\bibitem[{{Nash}(1990)}]{Nas90}
{Nash}, A.~G. 1990, Astrophys. J., Suppl. Ser., 72, 303

\bibitem[{{Ohishi} {et~al.}(1992){Ohishi}, {Irvine}, \& {Kaifu}}]{OhiIrv+92}
{Ohishi}, M., {Irvine}, W., \& {Kaifu}, N. 1992, in Astrochemistry of cosmic
  phenomena: proceedings of the 150th Symposium of the International
  Astronomical Union, held at Campos do Jordao, Sao Paulo, Brazil, August 5-9,
  1991. Dordrecht: Kluwer, ed. P.~D. {Singh}, 171

\bibitem[{{Pety}(2005)}]{pety05}
{Pety}, J. 2005, in SF2A-2005: Semaine de l'Astrophysique Francaise, ed.
  F.~{Casoli}, T.~{Contini}, J.~M. {Hameury}, \& L.~{Pagani}, 721

\bibitem[{{Price} {et~al.}(2003){Price}, {Viti}, \& {Williams}}]{PriVit+03}
{Price}, R.~J., {Viti}, S., \& {Williams}, D.~A. 2003, Mon. Not. R. Astron.
  Soc., 343, 1257

\bibitem[{{Rachford} {et~al.}(2002){Rachford}, {Snow}, {Tumlinson}, {Shull},
  {Blair}, {Ferlet}, {Friedman}, {Gry}, {Jenkins}, {Morton}, {Savage},
  {Sonnentrucker}, {Vidal-Madjar}, {Welty}, \& {York}}]{RacSno+02}
{Rachford}, B.~L., {Snow}, T.~P., {Tumlinson}, J., {Shull}, J.~M., {Blair},
  W.~P., {Ferlet}, R., {Friedman}, S.~D., {Gry}, C., {Jenkins}, E.~B.,
  {Morton}, D.~C., {Savage}, B.~D., {Sonnentrucker}, P., {Vidal-Madjar}, A.,
  {Welty}, D.~E., \& {York}, D.~G. 2002, ApJ, 577, 221

\bibitem[{{Smith} {et~al.}(2004){Smith}, {Pavlovski}, {MacLow}, {Rosen},
  {Khanzadyan}, {Gredel}, \& {Stanke}}]{SmiPav+04}
{Smith}, M.~D., {Pavlovski}, G., {MacLow}, M.-M., {Rosen}, A., {Khanzadyan},
  T., {Gredel}, R., \& {Stanke}, T. 2004, Astrophys. Space. Sci., 289, 333

\bibitem[{{Snow} \& {McCall}(2006)}]{SnoMcC06}
{Snow}, T.~P. \& {McCall}, B.~J. 2006, Ann. Rev. Astrophys. Astron., 44, 367

\bibitem[{{Viti} {et~al.}(2000){Viti}, {Williams}, \& {O'Neill}}]{VitWil+00}
{Viti}, S., {Williams}, D.~A., \& {O'Neill}, P.~T. 2000, A\&A, 354, 1062

\end{thebibliography}

\end{document}